\documentclass[11pt]{article}
\usepackage{cite}
\usepackage{graphicx}
\usepackage{indentfirst}
\usepackage{amsmath}
\usepackage{amssymb}
\usepackage{amsthm}
\usepackage{amsfonts}

\usepackage{anysize}
\usepackage{fancyhdr}
\title{An Energy-Based Comparison of Long-Hop and Short-Hop Routing in MIMO Networks}
\author{Caleb K. Lo, Sriram Vishwanath and Robert W. Heath, Jr. \\ Wireless Networking and Communications Group \\ Department of Electrical and Computer Engineering \\ The University of Texas at Austin \\ 1 University Station C0803 \\ Austin, TX 78712-0240 \\ Phone: (512) 471-1190 \\ Fax: (512) 471-6512 \\ Email: \{clo, sriram, rheath\}@ece.utexas.edu}
\date{}

\begin{document}

\maketitle

\begin{abstract}
This paper considers the problem of selecting either routes that consist of long hops or routes that consist of short hops in a network of multiple-antenna nodes, where each transmitting node employs spatial multiplexing.  This distance-dependent route selection problem is approached from the viewpoint of energy efficiency, where a route is selected with the objective of minimizing the transmission energy consumed while satisfying a target outage criterion at the final destination.  Deterministic line networks and two-dimensional random networks are considered.  It is shown that when 1) the number of hops traversed between the source and destination grows large or 2) when the target success probability approaches one or 3) when the number of transmit and/or receive antennas grows large, short-hop routing requires less energy than long-hop routing.  It is also shown that if both routing strategies are subject to the same delay constraint, long-hop routing requires less energy than short-hop routing as the target success probability approaches one.  In addition, numerical analysis indicates that given loose outage constraints, only a small number of transmit antennas are needed for short-hop routing to have its maximum advantage over long-hop routing, while given stringent outage constraints, the advantage of short-hop over long-hop routing always increases with additional transmit antennas.
\end{abstract}

Keywords - Random networks, large-antenna limits, multiple antennas, relays.

\section{Introduction}\label{intro}
By combining simple yet powerful signal processing techniques with informed node deployment, relays can significantly improve both signal quality and achieved data rates in next-generation wireless networks \cite{PabETAL:RelaBaseDeplConc:Sep:04}.  Many relay-based communication strategies have been proposed in the literature, including the direct-link three-terminal channel originally proposed by van der Meulen \cite{Meu:ThreTermCommChan:71} and the multi-source cooperative diversity strategy originally studied by Sendonaris et al \cite{SenETAL:UserCoopDiveSyst:Nov:03}.  In this paper, we are primarily concerned with multihop relaying where direct communication is not possible between a single source and its intended destination \cite{BoyETAL:MultDiveWireRela:Oct:04, DohETAL:CapaDistPhyLaye:Mar:06}.  Multihop relaying involves the source data being forwarded over several ``hops'' between intermediate relay nodes until reaching the destination.

One method for improving both multihop transmission, and relaying in general, is the use of multi-input multi-output (MIMO) signaling.  Recent work has shown that by exploiting the additional spatial degrees of freedom, MIMO relaying can yield the key benefits of improved signal quality and increased throughput \cite{FanTho:MIMOConfRelaChan:May:07, WanETAL:CapaMIMORelaChan:Jan:05, BolETAL:CapaScalLawsMIMO:Jun:06, TanHua:OptiDesiNonRege:Apr:07}.  The rewards of multiple-antenna relaying come at a price, though.  In particular, multiple-antenna signaling leads to increased per-node cost and complexity due to the need to deploy multiple RF chains at each node.  Also, the power drain required to operate the multiple RF chains may be problematic in a network of battery-powered nodes.

The transmission energy of MIMO relaying is closely connected to the lengths of the ``hops'' that are employed, motivating a study of the impact of routing hop length on energy consumption in a wireless network of multiple-antenna nodes.  The most related work to this paper is a study by Haenggi for single-antenna networks \cite{Hae:RoutRandRaylFadi:Jul:05}.  Multihop relaying in deterministic line networks and two-dimensional random networks is considered in \cite{Hae:RoutRandRaylFadi:Jul:05} with the objective of satisfying an outage constraint at the destination node.  While previous studies used unrealistic ``disk'' models for signal reception to conclude that short-hop routing consumed less energy than long-hop routing, the objective of \cite{Hae:RoutRandRaylFadi:Jul:05} is to perform this comparison subject to Rayleigh fading.  It is shown in \cite{Hae:RoutRandRaylFadi:Jul:05} that Rayleigh fading significantly closes the performance gap between short-hop and long-hop routing, and given appropriate delay constraints, long-hop routing actually consumes less energy than short-hop routing.

In this paper, we study the impact on transmit energy consumption of hop length-based routing in a MIMO network.  We employ the signal model in \cite{HocETAL:MultAnteChanHard:Sep:04}, where each transmitting node does not have channel state information and employs spatial multiplexing.  As in \cite{Hae:RoutRandRaylFadi:Jul:05}, we consider both deterministic and random networks with an outage constraint at the destination, where an outage occurs if the achieved mutual information between a transmitter-receiver pair falls below a target rate.  This mutual information expression is much more unwieldy than the analogous single-antenna expression in \cite{Hae:RoutRandRaylFadi:Jul:05}.  To obtain a more tractable expression for the mutual information, we employ the large-antenna limiting results in \cite{HocETAL:MultAnteChanHard:Sep:04} and approximate the mutual information as a Gaussian random variable.

We then use this Gaussian approximation to solve for the energy required to satisfy the target outage constraint at the destination.  We compare the required energy for both short-hop and long-hop routing in several limiting cases.  First, we show that 1) as the number of hops traversed between the source and the destination grows large or 2) as the target success probability approaches one or 3) as the number of transmitter and/or receiver antennas grows large, short-hop routing outperforms long-hop routing.  We then show that long-hop routing outperforms short-hop routing when the same delay constraint is applied to both approaches and the target success probability approaches one.

In our numerical analysis, we study the impact of increasing the number of transmit antennas on the energy ratio of short-hop to long-hop routing and observe a ``crossing point.''  Specifically, increasing the number of transmit antennas causes the ratio to increase for moderate values of the target success probability, implying that when more transmission failures can be tolerated, only a small number of antennas are needed to yield the maximum benefit for short-hop routing.  Also, increasing the number of transmit antennas causes the ratio to monotonically decrease for high values of the target success probability, implying that given stringent outage constraints, short-hop routing always benefits from additional transmit antennas.

We note that the problem of choosing either a long-hop route or a short-hop path is just another instance of the well-studied relay selection problem \cite{BleETAL:CoopCommOutaOpti:Sep:07, BerAdv:SeleCoopMultSour:Jan:08, KriETAL:AmplForwPartRela:Apr:08, YanETAL:NoveLocaRelaSele:Mar:08, MicKar:PhyLayeFairAmpl:Mar:08, SanETAL:BandEffiGeogMult:May:07, OecBoc:BidiRegeHalfDupl:May:08}.  Even though the relay selection problem is inherently difficult due to its dependence on multiple layers of the Open Systems Interconnection (OSI) protocol stack, we can glean valuable insights by focusing on a small subset of the key system parameters.  In this case, relays are selected by considering transmission energy and the inter-node distances in the network.

We also note that we evaluate energy consumption in random networks for two key reasons.  First, problems formulated on random networks are usually mathematically tractable, since the study of random networks has a solid theoretical foundation \cite{Gil:RandPlanNetw:Dec:61, StoETAL:StocGeomAppl:96}.  Second, the behavior of random networks can be used to model real-world networks, and the resulting insights can prove invaluable for network designers.  For example, if network connectivity is essential in a real-world network, the network designers can determine the conditions required for connectivity in a random network and consider these conditions when planning node deployments \cite{Vec:AdvaTeleNetw:07}.  The analytical and applied benefits of random network analysis have led to a flurry of stochastic geometry-inspired research \cite{RamETAL:CapaNetwCodiRand:Aug:05, MadETAL:ProdMultFlowWire:Apr:08, XueKum:ThetCoveConnLarg:Jun:06, AerSal:WireAdHocNetw:Jun:07, RadBou:OptiPoweContSche:Sep:04, Ros:MeanInteDistRegu:Aug:92}.

This paper is organized as follows.  In Section II we describe the system model for both network classes that we consider and present the routing strategies of interest.  We present a summary of our results on the energy-based comparison between long-hop and short-hop routing in Section III.  After presenting some numerical analysis and simulation results in Section IV, we conclude the paper in Section V.  The Appendix contains derivations of our key results from Section III.

The notation used throughout the paper is as follows.  We use boldface to denote matrices.  $\log(\cdot)$ denotes the base-2 logarithm.  $\textbf{I}_{N_t}$ denotes the $N_t\times N_t$ identity matrix.  $\textbf{A}^{\dagger}$ denotes the Hermitian conjugate of a matrix $\textbf{A}$.  $P(B)$ denotes the probability of the event $B$ occurring.  $\mathbb{E}$ denotes the expectation operator.

\section{System Model}\label{system-model}
\begin{figure}[tb]
\begin{center}
\includegraphics[width=3.0in]{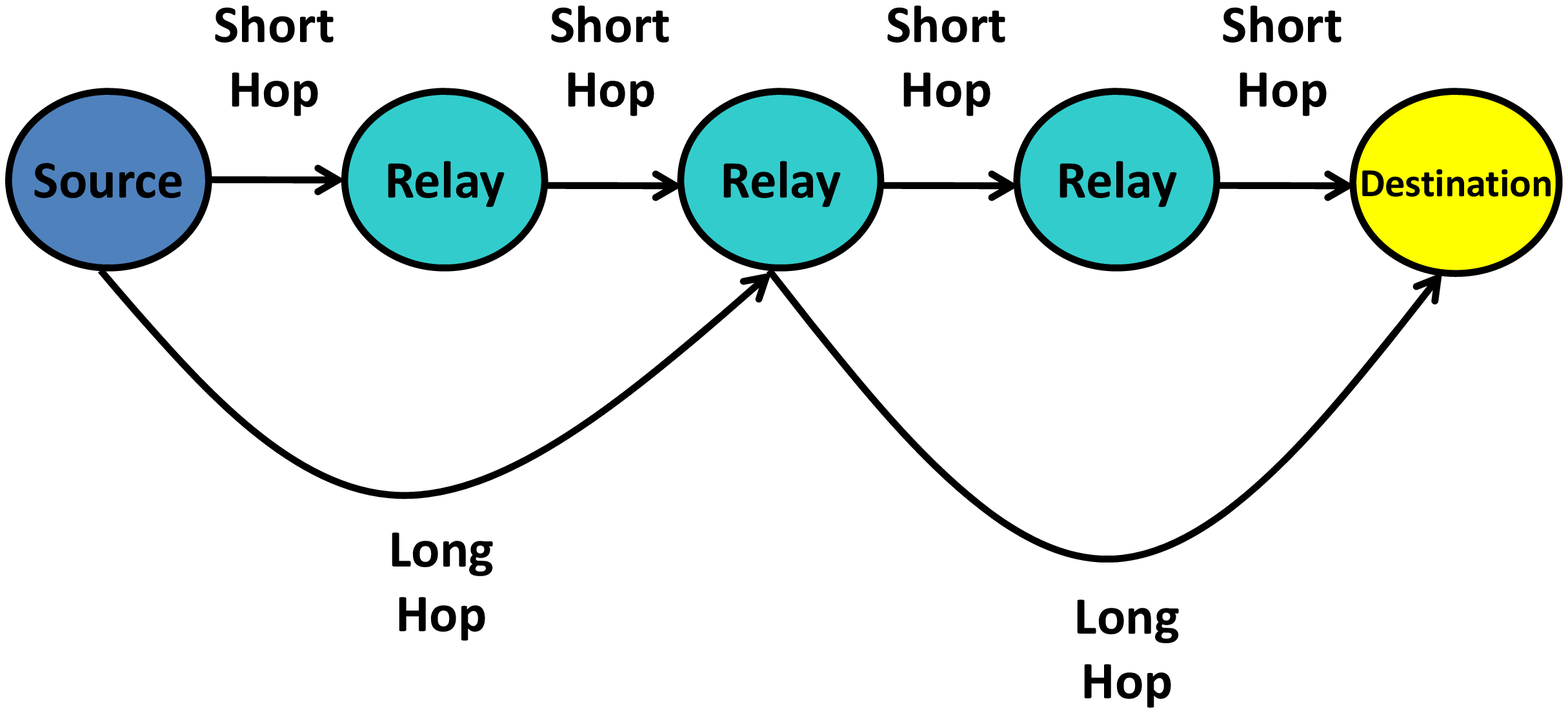}
\end{center}
\caption{Line network between source and destination with equidistant relays.}
\label{line-network}
\end{figure}

We consider two types of networks in this paper.  The first type of network that we consider is the deterministic line network model in Fig. \ref{line-network}.  Each neighboring pair of nodes in the network is separated by a fixed distance $d$.

For this network, the short-hop routing strategy is as follows.  The source node initially transmits a message to its nearest neighbor in the direction of the destination.  This nearest neighbor attempts to decode the source message, and if it succeeds, it retransmits the source message to its nearest neighbor in the direction of the destination.  The process continues until the destination receives the source message.

In contrast, the long-hop strategy involves the source node directly transmitting a message to its $n$-th nearest neighbor in the direction of the destination.  This $n$-th nearest neighbor attempts to decode the source message, and if it succeeds, it retransmits the source message to its $n$-th nearest neighbor in the direction of the destination.  The process continues until the destination receives the source message.  Note that the short-hop strategy requires $n$ transmissions for every transmission in the long-hop strategy.

\begin{figure}[tb]
\begin{center}
\includegraphics[width=3.0in]{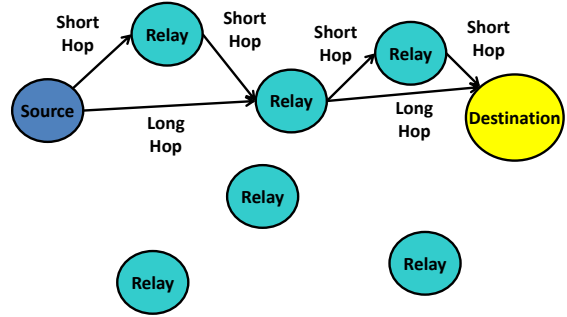}
\end{center}
\caption{Randomly dispersed nodes between source and destination.}
\label{random-network}
\end{figure}

The second type of network that we consider is the two-dimensional (2-D) random network model in Fig. \ref{random-network}.  Based on the exposition in \cite[Section 2.B]{Hae:RoutRandRaylFadi:Jul:05}, this network is generated via a Poisson point process of intensity $\lambda$ in the 2-D plane.  In particular, the probability of having $n$ nodes in a given area $A$ is
\begin{equation}
P(\textnormal{n nodes in A}) = e^{-\lambda A}\frac{(\lambda A)^n}{n!}.
\end{equation}
As in \cite{Hae:RoutRandRaylFadi:Jul:05} we set $\lambda = 1$ without loss of generality.  We employ the Poisson point process model for two reasons.  First, it corresponds to a uniform distribution when conditioning on the number of nodes to be dispersed in the 2-D plane.  Thus, it is an appropriate model for sensor networks that consist of randomly placed nodes, including battlefield sensors that are dropped during an airborne reconnaissance mission.  Second, the Poisson point process model facilitates the subsequent analysis in our paper.  Other point processes are more appropriate for modeling a wider class of ad hoc networks, though these models are relatively intractable and employing them requires a solid grasp of Palm theory \cite{StoETAL:StocGeomAppl:96}.  This is beyond the scope of our paper.

For this network, we adopt the short-hop routing of Strategy A from \cite[Section 2.D]{Hae:RoutRandRaylFadi:Jul:05}.  The notion of $\textit{progress}$ is essential to facilitate routing in the 2-D plane.  As defined in \cite[Section 2.B]{Hae:RoutRandRaylFadi:Jul:05}, $\textit{positive}$ progress in routing occurs when the $x$-distance between a selected node on the route and the destination decreases.  The basic idea behind Strategy A, then, is to consider the source-to-destination line and draw a sector of angle $\phi$ about it.  Next, the source node initially transmits a message to the closest node within this sector such that positive progress is made.  This nearest neighbor attempts to decode the source message, and if it succeeds, it retransmits the source message to the closest node within the sector such that positive progress is again made.  The process continues until the destination receives the source message.

We also adopt the long-hop routing of Strategy B from \cite[Section 2.D]{Hae:RoutRandRaylFadi:Jul:05}, where the source node directly transmits its message to the $n$-th node in the route of Strategy A.  This node attempts to decode the source message, and if it succeeds, it retransmits the source message to the $2n$-th node in the route of Strategy A.  The process continues until the destination receives the source message.  As in the case of the deterministic line network, Strategy A requires $n$ transmissions for every transmission in Strategy B.

Note that we do not consider the performance impact of amplify-and-forward relaying \cite{HasAlo:EndEndPerfTran:Nov:03,Kar:PerfBounMultWire:Mar:06} in this paper.  We constrain the relays to employ a decode-and-forward approach in order to build on the theoretical framework in \cite{Hae:RoutRandRaylFadi:Jul:05}.  We remark that the energy comparison in this paper can be performed in an amplify-and-forward network, where the objective for each routing strategy is to satisfy a target outage constraint at the destination.  It is evident that target outage criteria do not need to be satisfied at any of the relays in an amplify-and-forward network.  On one hand, if the amplify-and-forward relays have either full channel state information or knowledge of the channel statistics, they can possibly adjust their amplification factors to give short-hop routing an advantage over long-hop routing.  On the other hand, it should be stated that increased noise amplification results from employing additional hops, so it is not clear as to whether short-hops outperform long-hops in this case.

\subsection{Key Assumptions}
We make the following critical assumptions in this paper:
\begin{itemize}
\item In both the deterministic and the 2-D random networks, each node is equipped with $N_t$ transmit antennas and $N_r$ receive antennas.
\item As in \cite[Section 1.A]{HocETAL:MultAnteChanHard:Sep:04}, we assume that each transmitting node sends independent Gaussian signals with equal average power over each of its antennas. 
\item The power of each signal is chosen such that the average signal-to-noise ratio (SNR) at each receive antenna is $\rho$.  In particular, this average SNR value includes the effects of path loss.
\item The elements of each channel matrix $\textbf{H}_{i,a}$ between transmitting node $i$ and receiving node $a$ are circularly symmetric complex Gaussian zero-mean random variables, each with variance 0.5 for its real and imaginary parts.  This assumption simplifies our analysis and is typically used in the literature to obtain insights on the performance of real-world wireless systems.
\item Additive noise that consists of samples from a circularly symmetric complex Gaussian random process is present at each receiving node.  Each of the additive noise samples is a zero-mean random variable with variance 0.5 for its real and imaginary parts.
\item Consider a transmitting-receiving node pair $(i,a)$.  The receiving node $a$ has full knowledge of $\textbf{H}_{i,a}$, while the transmitting node $i$ only knows the distribution of the elements of $\textbf{H}_{i,a}$.  Note that allowing for limited feedback of $\textbf{H}_{i,a}$ from $a$ to $i$ could alter our results.
\item We do not consider the presence of interference, including external sources of RF energy, for the purposes of performance benchmarking.  As discussed in \cite[Section 2.A]{Hae:RoutRandRaylFadi:Jul:05}, the performance impact via changing the transmit energy is more apparent in a zero-interference network.  This is based on the fact that if the source node and all interferers identically scale their power, the signal-to-interference-plus-noise ratio (SINR) at any receiving node will only increase slightly.
\item We also do not consider the effects of lognormal shadowing for the purposes of analytical tractability.  It should be noted that for a network of single-antenna nodes, increasing the fading variance decreases the minimum node density that is required to obtain a connected network with high probability \cite{BetHar:ConnWireMultNetw:Sep:05}.  To the best of our knowledge, this result has not been extended to networks of multiple-antenna nodes.
\end{itemize}

\section{Energy Comparison}
In this section, we present our results on the energy required to transmit from the source node to the destination node in the networks in Figs. \ref{line-network} and \ref{random-network}.

Let $\textbf{H}$ be the channel between a neighboring pair of nodes.  Assuming an identity transmit covariance matrix, the mutual information between this node pair is \cite{HocETAL:MultAnteChanHard:Sep:04}
\begin{equation}\label{hochwald}
\mathcal{I} = \log\det\bigg(\textbf{I}_{N_t}+\frac{\rho}{N_t}\textbf{H}^{\dagger}\textbf{H}\bigg)
\end{equation}
where the transmitting node performs spatial multiplexing.  Note that the lack of channel state information at the transmitter precludes the use of transmission strategies such as antenna selection.

We want to achieve a success probability of $p$ between each neighboring node pair \cite{Hae:RoutRandRaylFadi:Jul:05}.  Thus, for a rate threshold of $R$
\begin{equation}
p = P(\mathcal{I} > R).
\end{equation}
We consider outage probability as the key performance metric in this paper for two reasons.  First, target data rates must be satisfied to enable efficient video transmission over commercial wireless networks such as HSPA \cite{HSPA}.  Second, this performance criterion facilitates our analytical contributions.  It should be clear, though, that the intuition gleaned from this information-theoretic metric cannot be directly applied to a practical system.  In particular, we are assuming that the transmitter employs Gaussian-distributed coding over infinite block lengths and that the receiver employs maximum-likelihood decoding.

\subsection{Gaussian Approximation}
We have
\begin{equation}
p = P\bigg(\log\det\bigg(\textbf{I}_{N_t}+\frac{\rho}{N_t}\textbf{H}^{\dagger}\textbf{H}\bigg) > R\bigg).
\end{equation}

It is proved in \cite[Theorem 1]{HocETAL:MultAnteChanHard:Sep:04} that, if we fix the number of transmit antennas $N_t$ and let the number of receive antennas $N_r$ grow large for each node, then the mutual information converges (in distribution) to a Gaussian random variable, and so
\begin{equation}
p \rightarrow \frac{1}{2}\textnormal{erfc}\bigg(\frac{R-N_t\log (1+\rho N_r/N_t)}{\sqrt{2N_t/N_r}\log(e)}\bigg). \label{solve-expr}
\end{equation}
As shown in \cite[Theorem 2]{HocETAL:MultAnteChanHard:Sep:04}, this result also holds if we fix the number of receive antennas $N_r$ and let the number of transmit antennas $N_t$ grow large for each node.  In addition, \cite[Theorem 3]{HocETAL:MultAnteChanHard:Sep:04} shows that this result holds if we let both $N_r$ and $N_t$ grow large and let either $\rho\rightarrow 0$ or $\rho\rightarrow\infty$.  Interestingly, the result in \cite[Theorem 3]{HocETAL:MultAnteChanHard:Sep:04} does not require $N_r/N_t$ to be fixed as both $N_r$ and $N_t$ grow large.

As illustrated by \cite[Fig. 1]{HocETAL:MultAnteChanHard:Sep:04}, approximating the mutual information as a Gaussian random variable is accurate even for $N_t = N_r = 2$ antennas.  Inspecting \cite[Fig. 1]{HocETAL:MultAnteChanHard:Sep:04} shows that the Gaussian approximation error is generally at most 5$\%$ when measuring the CDF of the mutual information.  As noted in \cite[Section 2.B]{HocETAL:MultAnteChanHard:Sep:04}, this approximation error decreases for larger values of $N_t$ and $N_r$.

Solving \eqref{solve-expr} for $\rho$ and letting $k = R/N_t-\sqrt{2/(N_tN_r)}(\log(e))\textnormal{erfc}^{-1}(2p)$ yields
\begin{equation}
\rho = \frac{N_t}{N_r}(2^{k}-1).
\end{equation}

\subsubsection{Deterministic Line Networks}\label{k-s-approx}
Since $\rho$ is the average SNR at each receive antenna, it follows that $\rho = E_0d^{-\alpha}/N_0$.  Assume that $p_r$ is the target success probability between the source and its $n$-th nearest neighbor in the direction of the destination.  Let $k_s = R/N_t-\sqrt{2/(N_tN_r)}(\log(e))\textnormal{erfc}^{-1}(2p_r)$.  The energy required to transmit over a single long-hop between the source and this $n$-th nearest neighbor is
\begin{equation}\label{line-long}
\begin{array}{lll}
E_s & = & N_0(n\cdot d)^{\alpha}\frac{N_t}{N_r}(2^{k_s}-1) \\
& = & n^{\alpha}\cdot (N_0d^{\alpha})\frac{N_t}{N_r}(2^{k_s}-1).
\end{array}
\end{equation}
Let $k_m = R/N_t-\sqrt{2/(N_tN_r)}(\log(e))\textnormal{erfc}^{-1}(2p_r^{1/n})$.  The energy required to transmit over $n$ short-hops between the source and this $n$-th nearest neighbor is
\begin{equation}\label{line-short}
E_m = n\cdot (N_0d^{\alpha})\frac{N_t}{N_r}(2^{k_m}-1).
\end{equation}
Thus, we want to compare $E_s$ and $E_m$ to determine the relative energy efficiency of long-hops and short-hops in a deterministic MIMO line network.

\subsubsection{2-D Random Networks}
Using the long-hop routing of Strategy B, the expected transmit energy, normalized by $N_0(N_t/N_r)$, can be derived from \cite[(29)]{Hae:RoutRandRaylFadi:Jul:05}
\begin{equation}
E_B = n^{\alpha}\bigg(\frac{2}{\phi}\bigg)^{\alpha/2}\Gamma\bigg(1+\frac{\alpha}{2}\bigg)\bigg(1-\frac{\alpha\phi^2(n-1)}{24n}\bigg)(2^{k_s}-1). \label{rand-long}
\end{equation}

Using the short-hop routing of Strategy A, the expected transmit energy, normalized by $N_0(N_t/N_r)$, can be derived from \cite[(24)]{Hae:RoutRandRaylFadi:Jul:05}
\begin{equation}
E_A = n\bigg(\frac{2}{\phi}\bigg)^{\alpha/2}\Gamma\bigg(1+\frac{\alpha}{2}\bigg)(2^{k_m}-1). \label{rand-short}
\end{equation}

Thus, we want to compare $E_B$ and $E_A$ to determine the relative energy efficiency of long-hops and short-hops in a random MIMO 2-D network.

\subsection{Limiting Cases for Line and 2-D Networks}\label{limit-line}
Now we state our results for the energy comparison between long-hop routing and short-hop routing in deterministic line networks and random 2-D networks.  First, we consider the limiting case where the number of hops $n$ for short-hop routing goes to infinity.

\newtheorem{thm-n-large}{Theorem}
\begin{thm-n-large}\label{thm-n-large}
Consider a deterministic line network.  Assuming that $\alpha > 1$ and $p_r\in (0.9,1)$,
\[\frac{E_m}{E_s}\rightarrow 0\]
as $n\rightarrow\infty$.  In other words, short-hop routing consumes less energy than long-hop routing to satisfy the same outage requirement in this regime.
\end{thm-n-large}

\begin{proof}
See Appendix \ref{proof-thm-1}.
\end{proof}

To interpret this result, recall the energy expressions in \eqref{line-long} and \eqref{line-short}.  As the hop-count $n$ increases, the energy consumed for long-hop routing scales as $n^{\alpha}$.  This posynomial scaling overcomes the impact of the monotonic decrease in the $\textnormal{erfc}^{-1}(\cdot)$ function on the energy consumed for short-hop routing.

Note that our assumption that $p_r\in (0.9,1)$ can be used to model typical end-to-end delivery requirements for Transmission Control Protocol (TCP) traffic and video content \cite{BalETAL:CompMechImprTCP:Dec:97}.  We now extend this result to 2-D networks as follows.

\newtheorem{cor-n-large}{Corollary}
\begin{cor-n-large}\label{cor-n-large}
Consider a random 2-D network.  Assuming that $\alpha > 1$ and $\alpha\phi^2 < 24$,
\[\frac{E_A}{E_B}\rightarrow 0\]
as $n\rightarrow\infty$.  In other words, Strategy A consumes less energy than Strategy B to satisfy the same outage requirement in this regime.
\end{cor-n-large}

\begin{proof}
As shown in Appendix \ref{proof-thm-1}, $n^{1-\alpha}\cdot (2^{k_m}-1)/(2^{k_s}-1)\rightarrow 0$ as $n\rightarrow\infty$.

Note that the ratio between the energy consumed by short-hop and long-hop routing in this case is
\begin{equation}
\frac{E_A}{E_B} = n^{1-\alpha}\cdot \frac{2^{k_m}-1}{2^{k_s}-1}\cdot \frac{1}{1-\frac{\alpha\phi^2(n-1)}{24n}}. \label{cor-n-large-eq}
\end{equation}

Also, $1/(1-\alpha\phi^2(n-1)/(24n))\rightarrow 1/(1-\alpha\phi^2/24)$ as $n\rightarrow\infty$, which is finite based on the assumptions of the corollary.  It then follows that $E_A/E_B\rightarrow 0$ as $n\rightarrow\infty$.
\end{proof}

The analysis for Corollary \ref{cor-n-large} is slightly different from that for Theorem \ref{thm-n-large}, since the energy consumed by long-hop routing now includes a term $(1-\alpha\phi^2(n-1)/(24n))$ due to its increased path efficiency.  This increased path efficiency cannot overcome the posynomial energy scaling of $n^{\alpha}$ for long-hop routing, though.

It should be stressed that Theorem \ref{thm-n-large} and Corollary \ref{cor-n-large} involve an unfair comparison between short-hop and long-hop routing, as short-hop routing is subject to a relatively loose delay constraint.  This is the critical factor behind the asymptotic success of short-hop routing.  Nevertheless, we believe that these results serve two important purposes.  First, they provide the first known energy-based comparison of short-hop and long-hop routing in a multiple-antenna wireless network, laying the foundation for further work in this area.  Second, the mathematical tools that are introduced in Appendix \ref{proof-thm-1} will be applied in Section \ref{subsubsec-delay}, where a more practical system with the same delay constraint on both routing strategies is studied.

Next, we consider the limiting case where the target success probability $p_r$ goes to one.

\newtheorem{thm-pr-unity}[thm-n-large]{Theorem}
\begin{thm-pr-unity}\label{thm-pr-unity}
Consider a deterministic line network.  Assuming that $n^{1-\alpha} < 1/2$,
\[\bigg(\lim_{p_r\rightarrow 1}\frac{E_m}{E_s}\bigg) < 1.\]
In other words, short-hop routing consumes less energy than long-hop routing to satisfy the same outage requirement in this regime.
\end{thm-pr-unity}

\begin{proof}
Recall $f_1(n,p_r)$ as defined in \eqref{proof-thm1-f1}.  As $p_r$ approaches one, we obtain the following limit
\begin{equation}
\lim_{p_r\rightarrow 1} f_1(n,p_r) = 1. \nonumber
\end{equation}

It follows that $E_m/E_s < f_2(n,p_r)$ such that $\lim_{p_r\rightarrow 1} f_2(n,p_r) = n^{1-\alpha}\cdot 2 < 1$, which follows from the assumptions of the theorem.  This establishes the theorem.
\end{proof}

Again, recall the energy expressions in \eqref{line-long} and \eqref{line-short}.  As $p_r$ approaches one, note that the difference between $p_r^{1/n}$ and $p_r$ for fixed $n$ decreases steadily.  This yields a corresponding decrease in the difference between $\textnormal{erfc}^{-1}(2p_r)$ and $\textnormal{erfc}^{-1}(2p_r^{1/n})$.  The key difference between short-hop routing and long-hop routing in this limiting case, then, is the posynomial impact of the path loss exponent $\alpha$ on long-hop routing.

We now extend this result to 2-D networks as follows.

\newtheorem{cor-pr-unity}[cor-n-large]{Corollary}
\begin{cor-pr-unity}\label{cor-pr-unity}
Consider a random 2-D network.  Assuming that $n^{1-\alpha}/(1-\alpha\phi^2(n-1)/(24n)) < 1/2$,
\[\bigg(\lim_{p_r\rightarrow 1}\frac{E_A}{E_B}\bigg) < 1\]
In other words, Strategy A consumes less energy than Strategy B to satisfy the same outage requirement in this regime.
\end{cor-pr-unity}

\begin{proof}
From the proof of Theorem \ref{thm-pr-unity}, it follows that in this case, $E_A/E_B < g_2(n,p_r)$ such that \\
$\lim_{p_r\rightarrow 1} g_2(n,p_r) = n^{1-\alpha}\cdot 2/(1-\alpha\phi^2 (n-1)/(24n))$.  The assumption of the corollary that $n^{1-\alpha}/(1-\alpha\phi^2 (n-1)/(24n)) < 1/2$ establishes the corollary.
\end{proof}

The critical difference between Corollary \ref{cor-pr-unity} and Theorem \ref{thm-pr-unity} is the requisite constraint on the parameters $n$, $\alpha$ and $\phi$.  By fixing the hop-count $n$ for the short-hop strategy, the increased path efficiency of long-hop routing improves its energy consumption.

\begin{figure}[tb]
\begin{center}
\includegraphics[width=3.5in]{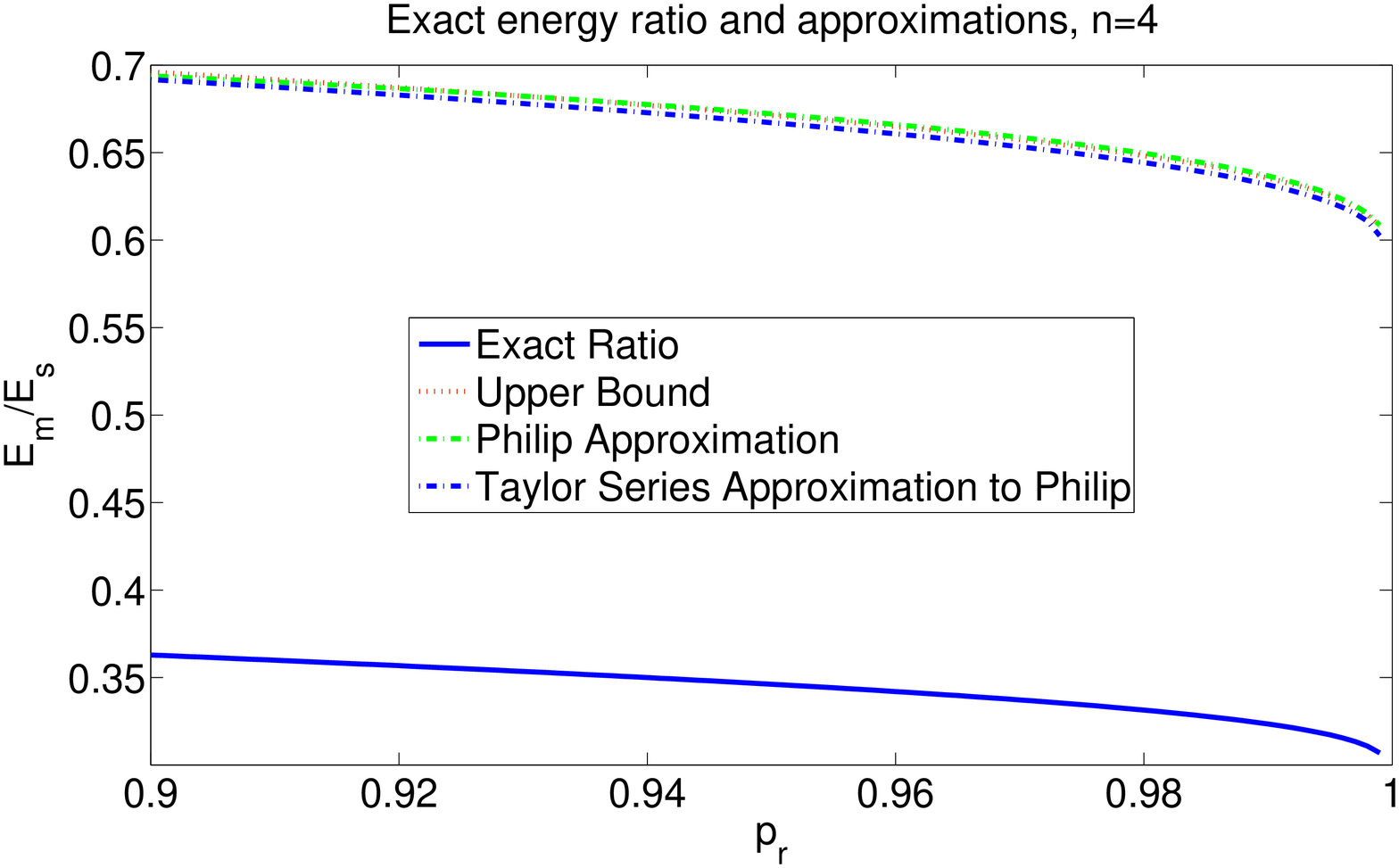}
\end{center}
\caption{Long-hop vs. short-hop energy for $n = 4$ hops.}
\label{sublinear-n4}
\end{figure}

It should be stressed that $E_m\rightarrow\infty$ as $p_r\rightarrow 1$, and the same is true for $E_s$, $E_A$ and $E_B$.  We believe that Theorem \ref{thm-pr-unity} and Corollary \ref{cor-pr-unity} serve an important purpose, though.  Specifically, these results provide valuable insights into the relative behavior of short-hop and long-hop routing for values of $p_r$ that are close to one.  Also, the behavior of the $\textnormal{erfc}^{-1}(x)$ function as $x\rightarrow 1$ is such that the transmit energy is moderate for these values of $p_r$.  These insights will be illustrated by Fig. \ref{sublinear-n4} in Example \ref{exp-line-network}.

Then, we consider the limiting case where the number of transmit antennas $N_t$ and/or the number of receive antennas $N_r$ goes to infinity.

\newtheorem{thm-ant-large}[thm-n-large]{Theorem}
\begin{thm-ant-large}\label{thm-ant-large}
Consider a deterministic line network.  Assuming that $n^{1-\alpha}\cdot 2 < 1$,
\[\frac{E_m}{E_s}\rightarrow 0\]
as $N_t\rightarrow\infty$ and/or $N_r\rightarrow\infty$.  In other words, short-hop routing consumes less energy than long-hop routing to satisfy the same outage requirement in this regime.
\end{thm-ant-large}

\begin{proof}
This result follows immediately from \eqref{em-es-bound}.
\end{proof}

In the large-antenna limit, which forms the basis of the results in \cite{HocETAL:MultAnteChanHard:Sep:04}, the path loss exponent $\alpha$ negatively impacts long-hop routing.

We now extend this result to 2-D networks as follows.

\newtheorem{cor-ant-large}[cor-n-large]{Corollary}
\begin{cor-ant-large}\label{cor-ant-large}
Consider a random 2-D network.  Assuming that $n^{1-\alpha}\cdot 2/(1-\alpha\phi^2(n-1)/(24n)) < 1$,
\[\frac{E_A}{E_B}\rightarrow 0\]
as $N_t\rightarrow\infty$ and/or $N_r\rightarrow\infty$.  In other words, Strategy A consumes less energy than Strategy B to satisfy the same outage requirement in this regime.
\end{cor-ant-large}

\begin{proof}
This result follows immediately from \eqref{cor-n-large-eq} and \eqref{em-es-bound}.
\end{proof}

As in Corollaries \ref{cor-n-large} and \ref{cor-pr-unity}, the increased path efficiency of long-hop routing narrows its energy gap with short-hop routing.

We now consider an example that further illustrates the energy efficiency of short-hop routing in a deterministic line network.

\newtheorem{exp-line-network}{Example}[section]
\begin{exp-line-network}\label{exp-line-network}
Line Network
\end{exp-line-network}

\begin{quote}
Consider a line network where $\alpha = 2$ and each node employs $N_t = N_r = 2$ antennas.  Assume that the short-hop strategy routes through $n = 3$ nearest neighbors.  Let $p_r\in (0.9,1)$.

As shown in Appendix \ref{proof-exp-1}, short-hop routing consumes less energy than long-hop routing.  Note that for $\alpha = 2$, short-hops yield the same energy consumed as long-hops in a network of single-antenna nodes \cite{Hae:RoutRandRaylFadi:Jul:05}.  When we consider a MIMO line network, it is apparent that using multiple-antenna transmission is inherently more energy-efficient than single-antenna transmission.  Specifically, MIMO requires less energy than single-input single-output (SISO) signaling to achieve the same target success probability.  Thus, MIMO short-hops should be even more energy-efficient than SISO short-hops for $\alpha = 2$.

Similar results can be obtained for the cases of $n = 4$ and $n = 5$ hops, and Fig. \ref{sublinear-n4} shows the long-hop versus short-hop energy comparison for $n = 4$ hops.  It is clear that even though the upper bound in \eqref{em-es-bound} is quite loose, we still observe that short-hop routing outperforms long-hop routing.  It is also evident that short-hops outperform long-hops under the approximations in steps (a) and (d) of \eqref{exp-line-network-array}.  Note that step (a) of \eqref{exp-line-network-array} is based on the approximation in \eqref{philip-approx}, and the tightness of this approximation to the upper bound in \eqref{em-es-bound} can be seen in Fig. \ref{sublinear-n4}.  In addition, as $p_r$ approaches one, short-hops continue to outperform long-hops, which illustrates our analytical result in Theorem \ref{thm-pr-unity} and shows that it can be extended to scenarios where finite transmit energy is expended.
\end{quote}

Next we consider an example that further illustrates the energy efficiency of short-hop routing in a random 2-D network.

\newtheorem{exp-2d-network}[exp-line-network]{Example}
\begin{exp-2d-network}\label{exp-2d-network}
2-D Network
\end{exp-2d-network}

\begin{quote}
Consider a 2-D network where $\alpha = 2$ and each node employs $N_t = N_r = 2$ antennas.  Assume that the short-hop Strategy A routes through $n = 3$ nearest neighbors in a sector with angle $\phi = \pi/2$.  Let $p_r\in (0.9,1)$ given a target rate $R = 4$.

As shown in Appendix \ref{proof-exp-2}, Strategy A consumes less energy than Strategy B.  A similar analysis can be carried out for $n = 4$ hops and $R\in\{4,8,16\}$ along with $n = 5$ hops and $R\in\{4,8,16\}$, and it can be shown that in these cases, Strategy A requires less energy than Strategy B.  The inherent energy efficiency of MIMO leads to this result.
\end{quote}

\subsubsection{Delay Constraints}\label{subsubsec-delay}
Now we consider the impact of imposing the same delay constraint on the long-hop and short-hop routing strategies.  From \cite[Section 4.A]{Hae:RoutRandRaylFadi:Jul:05}, we impose a delay constraint of $n$ time slots on both strategies.
As in \cite[Section 4.A]{Hae:RoutRandRaylFadi:Jul:05}, we also assume that the long-hop routing strategy can exploit this flexible delay constraint by transmitting to its $n$-th nearest neighbor in the direction of the destination in each time slot.  We refer to this modification of the standard long-hop approach as ``multi-transmit'' long-hop routing.  Since the target success probability at this $n$-th nearest neighbor after $n$ time slots is $p_r$ for short-hop and ``multi-transmit'' long-hop routing, the per-slot target success probability $p_{r,1}$ for the ``multi-transmit'' long-hop strategy is
\begin{equation}
p_{r,1} = 1 - (1 - p_r)^{1/n}. \nonumber
\end{equation}
This is obtained by noting that $(1-p_r) = (1-p_{r,1})^n$.

First we consider routing in a deterministic line network.  Let $E_{s,mult}$ denote the energy required to transmit over a single long-hop between the source and its $n$-th nearest neighbor, assuming that the source transmits during $n$ time slots.  Let $k_{s,mult} = R/N_t-\sqrt{2/(N_tN_r)}(\log(e))\textnormal{erfc}^{-1}(2p_{r,1})$.  From \eqref{line-long}, we see that
\begin{equation}\label{line-long-mult}
\begin{array}{lll}
E_{s,mult} & = & n\cdot n^{\alpha}\cdot (N_0d^{\alpha})\frac{N_t}{N_r}(2^{k_{s,mult}}-1) \\
& = & n^{\alpha + 1}\cdot (N_0d^{\alpha})\frac{N_t}{N_r}(2^{k_{s,mult}}-1).
\end{array}
\end{equation}
This follows from the fact that the source uses the same transmit energy in each time slot.

Then we consider routing in a 2-D random network.  Let $E_{B,mult}$ denote the expected transmit energy, normalized by $N_0(N_t/N_r)$, for the long-hop routing of Strategy B assuming that the source transmits during $n$ time slots.  Using \eqref{rand-long}, we see that
\begin{equation}\label{rand-long-mult}
\begin{array}{lll}
E_{B,mult} & = & n\cdot n^{\alpha}\Big(\frac{2}{\phi}\Big)^{\alpha/2}\Gamma\Big(1+\frac{\alpha}{2}\Big)\Big(1-\frac{\alpha\phi^2(n-1)}{24n}\Big)(2^{k_{s,mult}}-1) \\
& = & n^{\alpha + 1}\Big(\frac{2}{\phi}\Big)^{\alpha/2}\Gamma\Big(1+\frac{\alpha}{2}\Big)\Big(1-\frac{\alpha\phi^2(n-1)}{24n}\Big)(2^{k_{s,mult}}-1).
\end{array}
\end{equation} 
Again, this follows from the fact that the source uses the same transmit energy in each time slot.

Now we compare the performance of ``multi-transmit'' long-hop routing with that of short-hop routing.

\newtheorem{thm-mult-m}[thm-n-large]{Theorem}
\begin{thm-mult-m}\label{thm-mult-m}
For a deterministic line network,
\[\frac{E_{s,mult}}{E_m}\rightarrow 0\]
as $p_r\rightarrow 1$.  In other words, ``multi-transmit'' long-hop routing consumes less energy than short-hop routing to satisfy the same outage requirement.
\end{thm-mult-m}

\begin{proof}
The ratio of the energies consumed by ``multi-transmit'' long-hop routing and short-hop routing is
\begin{equation}
\begin{array}{lll}
\frac{E_{s,mult}}{E_m} & = & n^{\alpha}\cdot\frac{2^{k_{s,mult}}-1}{2^{k_m}-1} \\
& \stackrel{(a)}{\approx} & n^{\alpha}\cdot 2^{k_{s,mult}-k_m} \nonumber
\end{array}
\end{equation}
where step (a) holds as $p_r\rightarrow 1$.

Then we apply the approximation for $\textnormal{erfc}^{-1}(\cdot)$ in \eqref{philip-approx} to obtain
\begin{equation}\label{proof-thm6-f4}
\begin{array}{lll}
k_{s,mult}-k_m & = & \sqrt{\frac{2}{N_tN_r}}\log(e)(\textnormal{erfc}^{-1}(2(1-p_r)^{1/n})-\textnormal{erfc}^{-1}(2-2p_r^{1/n})) \\
& \approx & \sqrt{\frac{2}{N_tN_r}}\log(e)\Bigg(\sqrt{-\ln\Big(\sqrt{\pi}(2(1-p_r)^{1/n})\sqrt{-\ln(2(1-p_r)^{1/n})}\Big)} \\
& & -\sqrt{-\ln\bigg(\sqrt{\pi}(2-2p_r^{1/n})\sqrt{-\ln(2-2p_r^{1/n})}\bigg)}\Bigg) \\
& \triangleq & f_4(n,p_r).
\end{array}
\end{equation}

In particular, we obtain the following limit
\begin{equation}
\lim_{p_r\rightarrow 1}n^{\alpha}\cdot 2^{f_4(n,p_r)} = 0. \nonumber
\end{equation}

It follows that $\lim_{p_r\rightarrow 1}E_{s,mult}/E_m = 0$, which establishes the theorem.
\end{proof}

As will be in seen in Section \ref{sim-res}, this result only holds for $p_r \approx 1$.  For smaller values of $p_r$, short-hop routing outperforms ``multi-transmit'' long-hop routing.

We now extend this result to 2-D networks as follows.

\newtheorem{cor-mult-m}[cor-n-large]{Corollary}
\begin{cor-mult-m}\label{cor-mult-m}
For a random 2-D network,
\[\frac{E_{B,mult}}{E_A}\rightarrow 0\]
as $p_r\rightarrow 1$.  In other words, ``multi-transmit'' Strategy B consumes less energy than Strategy A to satisfy the same outage requirement.
\end{cor-mult-m}

\begin{proof}
This follows directly from the proof of Theorem \ref{thm-mult-m}.
\end{proof}

Note that the asymptotic behavior of $p_r$ does not affect the path efficiency of ``multi-transmit'' Strategy B.  Also, as will be seen in Section \ref{sim-res}, this result only holds for $p_r \approx 1$.  For smaller values of $p_r$, Strategy A outperforms ``multi-transmit'' Strategy B.

\section{Numerical Analysis and Simulation Results}\label{sim-res}
Here we perform further energy-based comparisons of the long-hop and short-hop routing strategies.

\begin{figure}[tb]
\begin{center}
\includegraphics[width=3.5in]{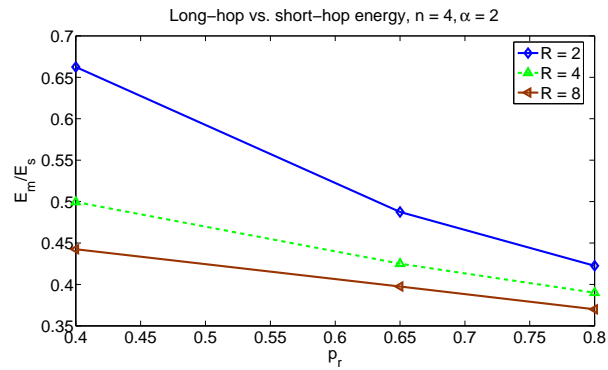}
\end{center}
\caption{Impact of target rate on energy comparison.}
\label{rate-effect}
\end{figure}

Fig. \ref{rate-effect} shows the impact of the target rate $R$ on the energy comparison between short-hops and long-hops.  We consider transmission in a deterministic line network with $n = 4$ hops and $N_t = N_r = 2$ transmit and receive antennas.  In addition, we fix $\alpha = 2$.

It is apparent that as the target end-to-end success probability $p_r$ increases, short-hop routing outperforms long-hop routing for all considered target rates.  Also, we observe that the energy advantage of short-hop routing over long-hop routing increases as the target rate $R$ increases.  It should be stressed that the benefits of short-hop routing for large target rates do not necessarily translate to an interference-limited environment.  As shown in \cite{SikETAL:BandPoweEffiRout:Jun:06} for an interference-limited network of single-antenna nodes, long-hop routing requires less energy than short-hop routing for sufficiently large target rates.

\begin{figure}[tb]
\begin{center}
\includegraphics[width=3.5in]{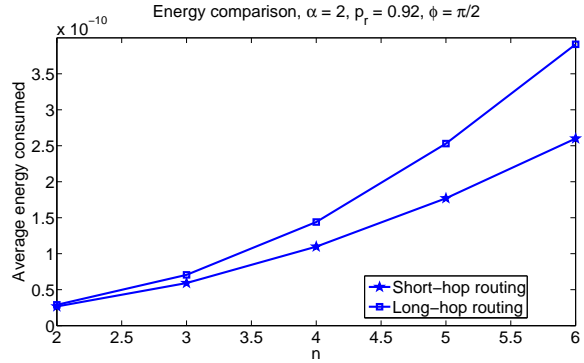}
\end{center}
\caption{Energy consumption in a 2-D random network.}
\label{energy-ppp}
\end{figure}

Fig. \ref{energy-ppp} compares the energy efficiency of long-hop and short-hop routing in a random 2-D network.  We uniformly distribute 30 points in a sector of angle $\phi = \pi/2$ between the source and the destination.  We also consider $\alpha = 2$ and set $p_r = 0.92$.  In addition, we set $N_t = N_r = 2$ along with $R = 2$.

We observe that short-hops consume less energy than long-hops as the hop-count $n$ increases, which illustrates our result in Corollary \ref{cor-n-large}.  Note that the energy advantage of short-hop routing increases as the hop-count increases.  From additional Monte Carlo simulation we determined that the energy advantage of short-hop routing does not change as the number of nodes in the network increases.  In addition, the energy consumed for long-hops increases with the hop-count due to longer distances traversed and the effects of path loss.

\begin{figure}[tb]
\begin{center}
\includegraphics[width=3.5in]{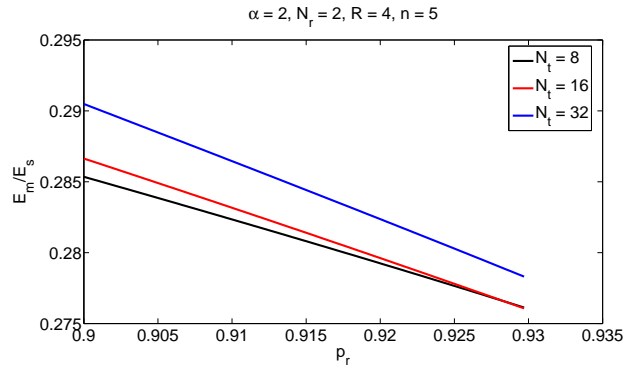}
\end{center}
\caption{Impact of $N_t$ on long-hop vs. short-hop energy given a loose outage constraint.}
\label{loose-qos}
\end{figure}

\begin{figure}[tb]
\begin{center}
\includegraphics[width=3.5in]{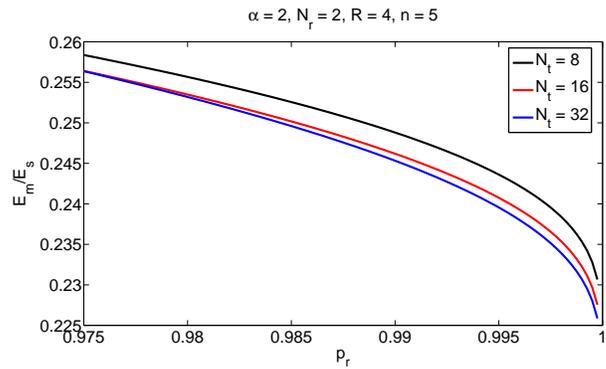}
\end{center}
\caption{Impact of $N_t$ on long-hop vs. short-hop energy given a strict outage constraint.}
\label{strict-qos}
\end{figure}

Fig. \ref{loose-qos} and Fig. \ref{strict-qos} illustrate the performance impact of $N_t$ and $p_r$ in a line network.  We consider $\alpha = 2$ and set $N_r = 2$.  We also set $R = 4$ for transmission over $n = 5$ hops with $p_r\in (0.9,1)$.

First, as shown in Fig. \ref{strict-qos}, adding transmit antennas causes the short-hop-to-long-hop energy ratio to monotonically decrease for $p_r\geq 0.98$.  This can be explained by noting that $k_m$ is dominated by the expression containing erfc$^{-1}(\cdot)$ for increasing $N_t$, and the same is true for $k_s$.  Also, the difference between the long-hop target $p_r$ and the per-hop short-hop target $p_r^{1/n}$ is small, implying that the difference between erfc$^{-1}(2p_r)$ and erfc$^{-1}(2p_r^{1/n})$ for long-hop and short-hop routing, respectively is small.  Thus, the path loss in \eqref{line-long} is the key factor that penalizes long-hop routing for stringent outage constraints, implying that short-hop routing always benefits from increasing $N_t$ in this regime.

Second, as shown in Fig. \ref{loose-qos}, adding transmit antennas eventually causes the energy ratio to increase for $p_r\leq 0.93$.  This behavior for looser outage constraints can be explained as in the previous paragraph.  In particular, the larger difference between the long-hop target $p_r$ and the relatively stricter per-hop short-hop target $p_r^{1/n}$ partially mitigates the advantages of short-hop routing for looser outage constraints, implying that short-hop routing derives its maximum benefit from a small number of transmit antennas.

\begin{figure}[tb]
\begin{center}
\includegraphics[width=3.5in]{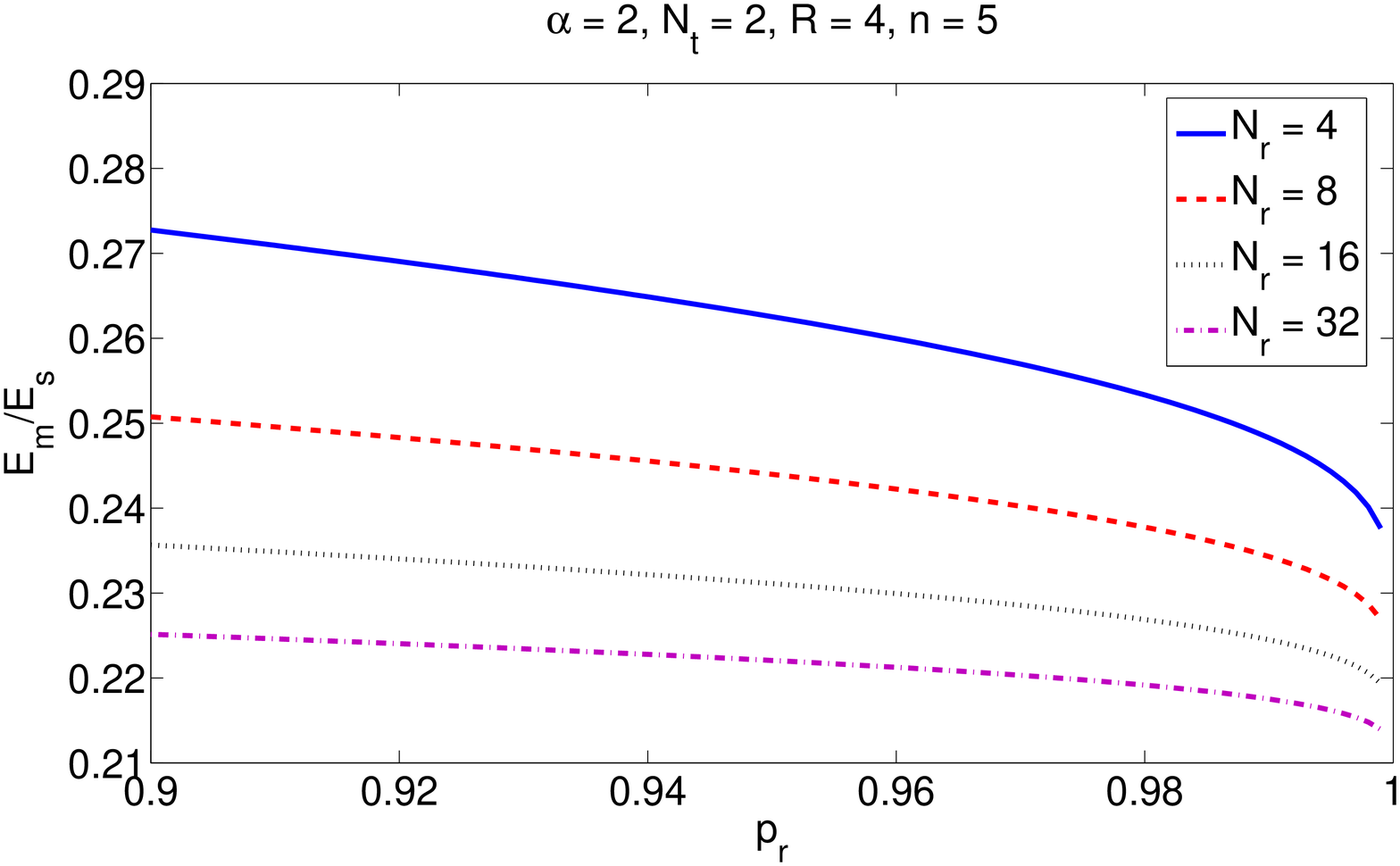}
\end{center}
\caption{Impact of $N_r$ on long-hop vs. short-hop energy.}
\label{rx-antennas}
\end{figure}

Fig. \ref{rx-antennas} shows the impact of $N_r$ on the energy ratio of short-hop to long-hop routing in a line network.  We employ most of the same parameters in Figs. \ref{loose-qos} and \ref{strict-qos}, except that we fix $N_t = 2$.

We observe that adding receive antennas causes the energy ratio to monotonically decrease for all $p_r\in (0.9,1)$.  This can be explained by inspecting \eqref{line-long} and \eqref{line-short}.  By fixing $N_t$ and increasing $N_r$, the performance impact of the outage constraint $p_r$ is minimized.  Then, the path loss in \eqref{line-long} becomes the key factor that penalizes long-hop routing, implying that short-hop routing always benefits from increasing $N_r$.

\begin{figure}[tb]
\begin{center}
\includegraphics[width=3.5in]{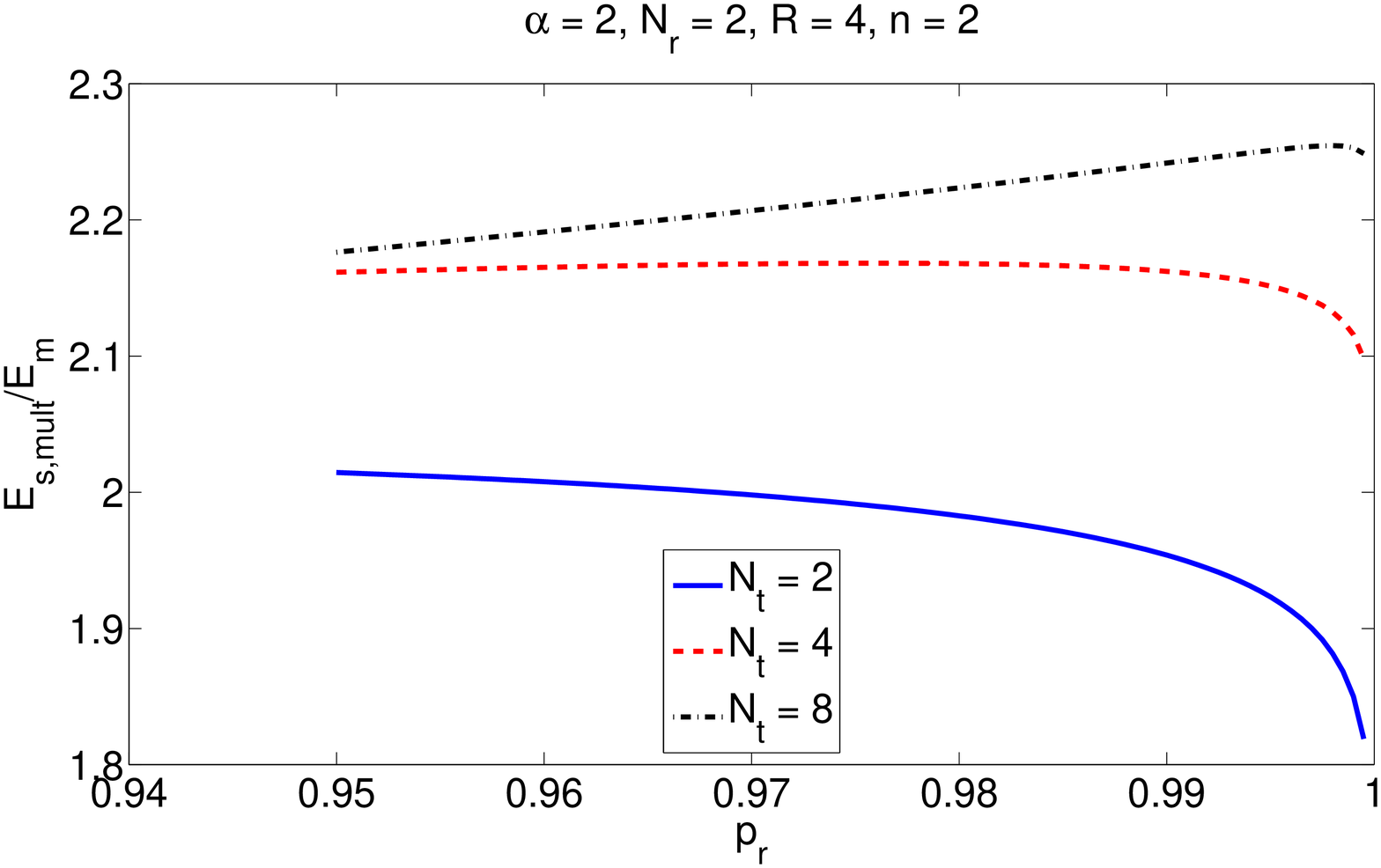}
\end{center}
\caption{Impact of $N_t$ on ``multi-transmit'' long-hop vs. short-hop energy for $N_r = 2$.}
\label{mult-short-line}
\end{figure}

Now we consider the impact of delay constraints.  Fig. \ref{mult-short-line} shows the impact of $N_t$ on the energy ratio of ``multi-transmit'' long-hop to short-hop routing in a line network.  We employ most of the same parameters in Figs. \ref{loose-qos} and \ref{strict-qos}, except that we set $p_r\in (0.95,1)$ and $n = 2$.

We observe that ``multi-transmit'' long-hop routing is outperformed by short-hop routing in this regime, and we are not able to illustrate the result in Theorem \ref{thm-mult-m} due to the finite precision arithmetic employed by Matlab.  Also, for a large number of transmit antennas $N_t$, short-hop routing gains an energy advantage over ``multi-transmit'' long-hop routing as $p_r$ increases.  This is due to the fact that as $p_r$ increases, the difference between the per-slot target success probabilities $p_r^{1/n}$ and $p_{r,1}$ for short-hop and ``multi-transmit'' long-hop routing, respectively, decreases.  Thus, ``multi-transmit'' long-hop routing is hindered by the source having to transmit in each time slot.

\begin{figure}[tb]
\begin{center}
\includegraphics[width=3.5in]{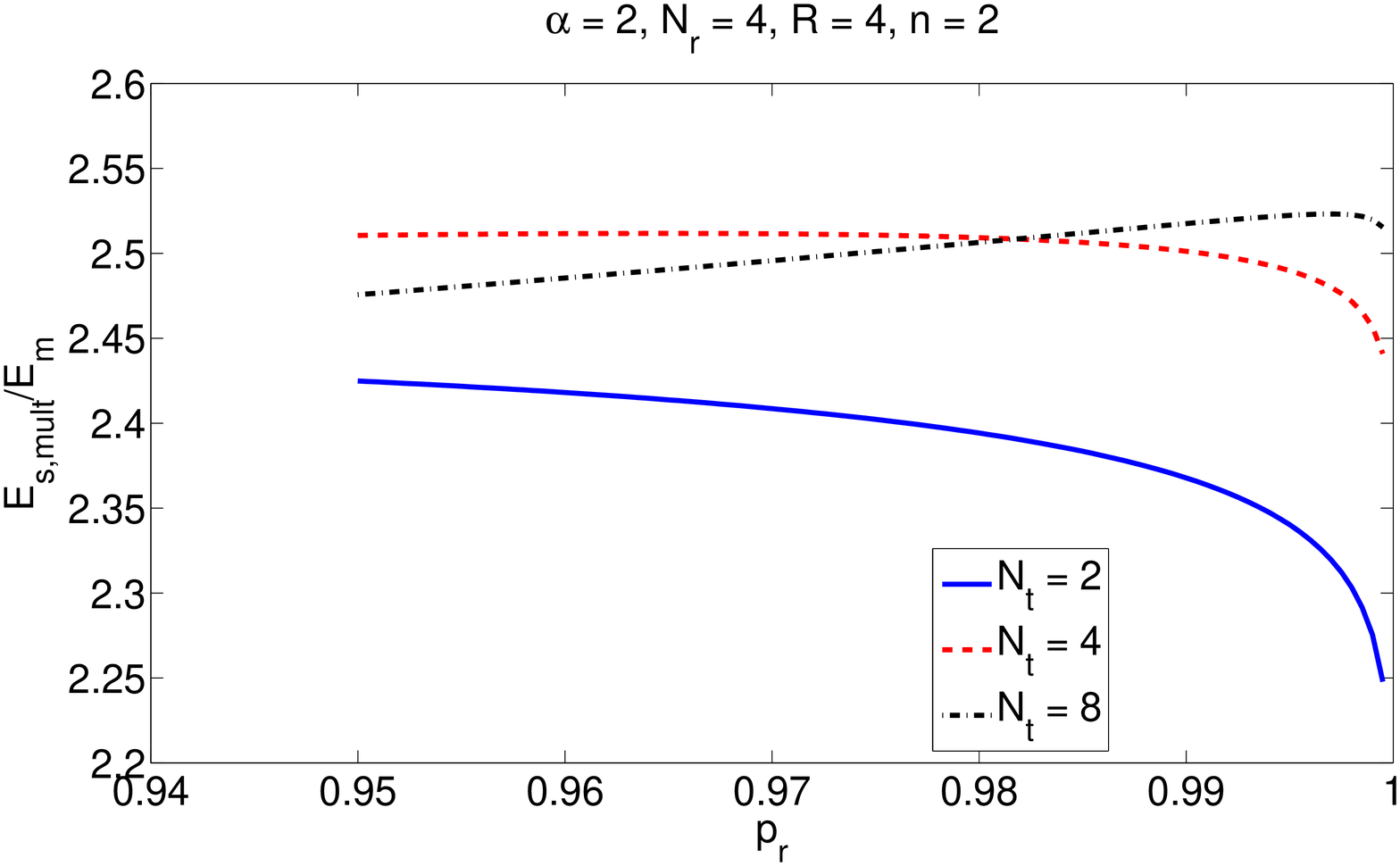}
\end{center}
\caption{Impact of $N_t$ on ``multi-transmit'' long-hop vs. short-hop energy for $N_r = 4$.}
\label{mult-short-line2}
\end{figure}

Fig. \ref{mult-short-line2} performs the same comparison as in Fig. \ref{mult-short-line} except that we set $N_r = 4$.  Interestingly, as in Fig. \ref{loose-qos}, we observe that short-hop routing begins to lose its energy advantage over ``multi-transmit'' long-hop routing as the number of transmit antennas $N_t$ increases beyond a certain level.  This behavior is observed for relatively low values of the target success probability $p_r$, and so the larger difference between $p_r$ and the per-hop short-hop target $p_r^{1/n}$ in this regime partially mitigates the energy advantage of short-hop routing.

\begin{figure}[tb]
\begin{center}
\includegraphics[width=3.5in]{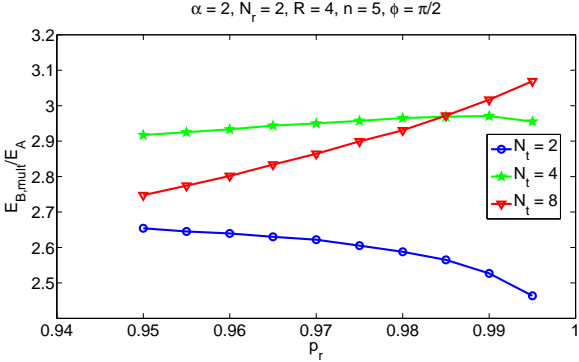}
\end{center}
\caption{Impact of $N_t$ on ``multi-transmit'' long-hop vs. short-hop energy in a 2-D random network.}
\label{mult-short-rand}
\end{figure}

Fig. \ref{mult-short-rand} performs the same comparison as in Figs. \ref{mult-short-line} and \ref{mult-short-line2}, except in a 2-D random network.  We employ the same parameters as in Fig. \ref{mult-short-line}, except that we set $\phi = \pi/2$ and $n = 5$.  In this case we observe analogous behavior to that in Fig. \ref{mult-short-line2}.

\section{Conclusion}
We have compared the performance of long-hop and short-hop routing strategies in MIMO networks in terms of energy consumption.  For both deterministic line networks and two-dimensional random networks, we have shown that short-hop routing actually improves upon long-hop routing in several limiting cases.  Our numerical analysis indicates that given loose outage constraints, only a small number of transmit antennas are needed for short-hop routing to have its maximum benefit, while given stringent outage constraints, short-hop routing always benefits from additional transmit antennas.  The obtained results imply that in MIMO systems with reasonably loose delay constraints, short-hop routing is a viable strategy for consideration.

The results in this paper should not be taken to conclude that short-hop routing is always preferable to long-hop routing in a MIMO network.  In particular, the benefits of long-hop routing, at least for single-antenna networks, are most fully realized when reasonably loose delay constraints are applied in conjunction with full channel state information at each transmitting node \cite{Hae:RoutRandRaylFadi:Jul:05}.  A more complete treatment of this problem, then, would consider the combined impact of these two factors on energy efficiency.  Also, while the Poisson point process is a useful model for random sensor node deployments, other wireless networks of interest may exhibit a more structured pattern.  As mentioned in Section \ref{system-model}, additional tools from stochastic geometry will be required to model the long-hop/short-hop comparison in such networks.

\appendix

\section{Proof of Theorem \ref{thm-n-large}}\label{proof-thm-1}
The ratio of the energies consumed by short-hop routing and by long-hop routing is
\begin{equation}\label{em-es-bound}
\begin{array}{lll}
\frac{E_m}{E_s} & = & n^{1-\alpha}\cdot\frac{2^{k_m}-1}{2^{k_s}-1} \\
& < & n^{1-\alpha}\cdot\frac{2^{k_m}}{2^{k_s}-1} \\
& < & n^{1-\alpha}\cdot 2^{k_m-k_s+1}.
\end{array}
\end{equation}
In particular, $k_m-k_s+1 = \sqrt{2/(N_tN_r)}\log(e)(\textnormal{erfc}^{-1}(2p_r)-\textnormal{erfc}^{-1}(2p_r^{1/n}))+1$.

Next we employ an approximation for $\textnormal{erfc}^{-1}$ due to Philip \cite{Phi:FuncInveThet:60}
\begin{equation}\label{philip-approx}
\textnormal{erfc}^{-1}(x) \approx \sqrt{-\ln\Big(\sqrt{\pi}x\sqrt{-\ln(x)}\Big)}.
\end{equation}
This approximation is particularly good for small $x$, and the error is less than $1.35\%$ for $x\leq 0.2$.  Note that we have assumed that $p_r\in (0.9,1)$, so to apply this approximation we need the relation $\textnormal{erfc}^{-1}(2-x) = -\textnormal{erfc}^{-1}(x)$ \cite[(1.6)]{Phi:FuncInveThet:60}.  Now we can apply \eqref{philip-approx} to yield
\begin{equation}\label{proof-thm1-f1}
\begin{array}{lll}
k_m-k_s+1 & = & \sqrt{\frac{2}{N_tN_r}}\log(e)(\textnormal{erfc}^{-1}(2-2p_r^{1/n})-\textnormal{erfc}^{-1}(2-2p_r))+1 \\
& \approx & \sqrt{\frac{2}{N_tN_r}}\log(e)\Bigg(\sqrt{-\ln\bigg(\sqrt{\pi}(2-2p_r^{1/n})\sqrt{-\ln(2-2p_r^{1/n})}\bigg)} \\
& & -\sqrt{-\ln\bigg(\sqrt{\pi}(2-2p_r)\sqrt{-\ln(2-2p_r)}\bigg)}\Bigg)+1 \\
& \triangleq & f_1(n,p_r).
\end{array}
\end{equation}
In particular, as $n$ grows large, we obtain the following limit
\begin{equation}
\lim_{n\rightarrow\infty}n^{1-\alpha}\cdot 2^{f_1(n,p_r)} = 0. \nonumber
\end{equation}

This does not immediately establish that $\lim_{n\rightarrow\infty}n^{1-\alpha}\cdot 2^{k_m-k_s+1} = 0$.  Now let
\begin{equation}
\begin{array}{lll}
a(n) & = & \sqrt{2/(N_tN_r)}\log(e)\textnormal{erfc}^{-1}(2-2p_r^{1/n}) \\
b(n) & = & \sqrt{2/(N_tN_r)}\log(e)\sqrt{-\ln\bigg(\sqrt{\pi}(2-2p_r^{1/n})\sqrt{-\ln(2-2p_r^{1/n})}\bigg)}. \nonumber
\end{array}
\end{equation}

We now know that
\begin{equation}
\lim_{n\rightarrow\infty}n^{1-\alpha}\cdot 2^{b(n)} = 0 \nonumber
\end{equation}
and so $\forall\epsilon_1 > 0,\exists N_1$ such that $|n^{1-\alpha}\cdot 2^{b(n)}| < \epsilon_1\quad\forall n > N_1$.

Also, we can write $a(n) = b(n)+c(n)$.  The key step is to observe from \cite[Table 2]{Phi:FuncInveThet:60} that
\begin{equation}
\lim_{n\rightarrow\infty}c(n) = 0 \nonumber
\end{equation}
since $2-2p_r^{1/n}\rightarrow 0$ as $n\rightarrow\infty$, and so $\forall\epsilon_2 > 0,\exists N_2$ such that $|c(n)| < \epsilon_2\quad\forall n > N_2$.

Let $N_3 = \max(N_1,N_2)$.  Then, $\forall n > N_3$, we have
\begin{equation}
\begin{array}{lll}
|n^{1-\alpha}\cdot 2^{a(n)}| & = & |n^{1-\alpha}\cdot 2^{b(n)+c(n)}| \\
& = & |n^{1-\alpha}\cdot 2^{b(n)}|\cdot |2^{c(n)}| \\
& < & \epsilon_1\cdot 2^{\epsilon_2}. \nonumber
\end{array}
\end{equation}
This shows that $\forall\epsilon_3 > 0,\exists N_3$ such that $|n^{1-\alpha}\cdot 2^{a(n)}| < \epsilon_3\quad\forall n > N_3$, where $\epsilon_3 = \epsilon_1\cdot 2^{\epsilon_2}$, and so we have
\begin{equation}
\lim_{n\rightarrow\infty}n^{1-\alpha}\cdot 2^{a(n)} = 0. \nonumber
\end{equation}

It now follows from \eqref{em-es-bound} that $\lim_{n\rightarrow\infty}E_m/E_s = 0$, which establishes Theorem \ref{thm-n-large}. 

\section{Computations for Example \ref{exp-line-network}}\label{proof-exp-1}
Consider a deterministic line network with the parameters from Example \ref{exp-line-network}.  We see that for short-hops to be more energy-efficient than long-hops, we need $(2^{k_m}-1)/(2^{k_s}-1) < 3$.  By employing the same approach used to obtain \eqref{em-es-bound}, we want to show that $k_m-k_s+1 < \log(3)$.

Now we apply the same approximation due to Philip from Appendix \ref{proof-thm-1} and obtain
\begin{equation}\label{exp-line-network-array}
\begin{array}{lll}
f(p_r) & \triangleq & k_m-k_s+1 \\
& = & 1.02(\textnormal{erfc}^{-1}(2-2p_r^{1/3})-\textnormal{erfc}^{-1}(2-2p_r))+1 \\
& \stackrel{(a)}{\approx} & 1.02\sqrt{-\ln(2\sqrt{\pi})-\ln(1-p_r^{1/3})-\frac{1}{2}\ln(-\ln(2)-\ln(1-p_r^{1/3}))} \\
& & -1.02\sqrt{-\ln(2\sqrt{\pi})-\ln(1-p_r)-\frac{1}{2}\ln(-\ln(2)-\ln(1-p_r))}+1 \\
& \stackrel{(b)}{\approx} & 1.02\sqrt{-\ln(2\sqrt{\pi})+\sum_{k=1}^{10^4}\frac{1}{k}p_r^{k/3}-\frac{1}{2}\ln(-\ln(2)-\ln(1-p_r^{1/3}))} \\
& & -1.02\sqrt{-\ln(2\sqrt{\pi})+\sum_{k=1}^{10^4}\frac{1}{k}p_r^k-\frac{1}{2}\ln(-\ln(2)-\ln(1-p_r))}+1 \\
& \stackrel{(c)}{\approx} & 1.02\sqrt{-\ln(2\sqrt{\pi})+\sum_{k=1}^{10^4}\frac{1}{k}p_r^{k/3}-\frac{1}{2}\ln(-\ln(2)+p_r^{1/3})} \\
& & -1.02\sqrt{-\ln(2\sqrt{\pi})+\sum_{k=1}^{10^4}\frac{1}{k}p_r^k-\frac{1}{2}\ln(-\ln(2)+p_r)}+1 \\
& \stackrel{(d)}{\approx} & 1.02\sqrt{-\ln(2\sqrt{\pi})+\sum_{k=1}^{10^4}\frac{1}{k}p_r^{k/3}-\frac{1}{2}(-1-\ln(2)+p_r^{1/3})} \\
& & -1.02\sqrt{-\ln(2\sqrt{\pi})+\sum_{k=1}^{10^4}\frac{1}{k}p_r^k-\frac{1}{2}(-1-\ln(2)+p_r)}+1 \\
& \triangleq & g(p_r)
\end{array}
\end{equation}
where we applied a Taylor series approximation with $10^4$ terms for $\ln(1-x)$ to obtain step (b) of \eqref{exp-line-network-array}, and we applied a first-order Taylor series approximation to obtain steps (c) and (d) of \eqref{exp-line-network-array}.

Our objective is to show that the approximation to $f(p_r)$ is monotone decreasing for $p_r\in (0.9,1)$.  We differentiate this approximation and observe its behavior in $p_r\in (0.9,1)$
\begin{equation}\label{diff-approx}
\begin{array}{lll}
g^{'}(p_r) & = & 0.51\cdot\frac{(1/6)p_r^{-2/3}+(1/3)\sum_{k=2}^{10^4}p_r^{k/3-1}}{\sqrt{-(1/2)\ln(2\pi)+(1/2)(1+p_r^{1/3})+\sum_{k=2}^{10^4}(1/k)p_r^{k/3}}} \\
& & -0.51\cdot\frac{(1/2)+\sum_{k=2}^{10^4}p_r^{k-1}}{\sqrt{-(1/2)\ln(2\pi)+(1/2)(1+p_r)+\sum_{k=2}^{10^4}(1/k)p_r^k}}.
\end{array}
\end{equation}
Note that since $x < x^{1/3}$ for $x\in (0.9,1)$, the denominator of the first term in \eqref{diff-approx} is greater than that of the second term in \eqref{diff-approx}.  Also, it is straightforward to show that $(1/6)p_r^{-2/3} < 1/2$ for $p_r\in (0.9,1)$.

Thus, to show that the approximation to $f(p_r)$ is monotone decreasing for $p_r\in (0.9,1)$ we need to show that $(1/3)p_r^{k/3-1} < p_r^{k-1}$ for $k\in\{2,3,\ldots,10^4\}$.  This is equivalent to showing that $(1/3)p_r^{k/3} < p_r$, which is clear for $k\geq 3$ since $p_r^m < p_r$ for $m > 1$ and $p_r < 1$.  As for $k = 2$, we must show that $(1/3)p_r^{2/3} < p_r$, which is equivalent to showing that $(1/3) < p_r^{1/3}$.  Since $0.9^{1/3} > (1/3)$ and $p_r^{1/3}$ is monotone increasing for $p_r\in (0.9,1)$, we have established this claim.

\begin{figure}[tb]
\begin{center}
\includegraphics[width=3.5in]{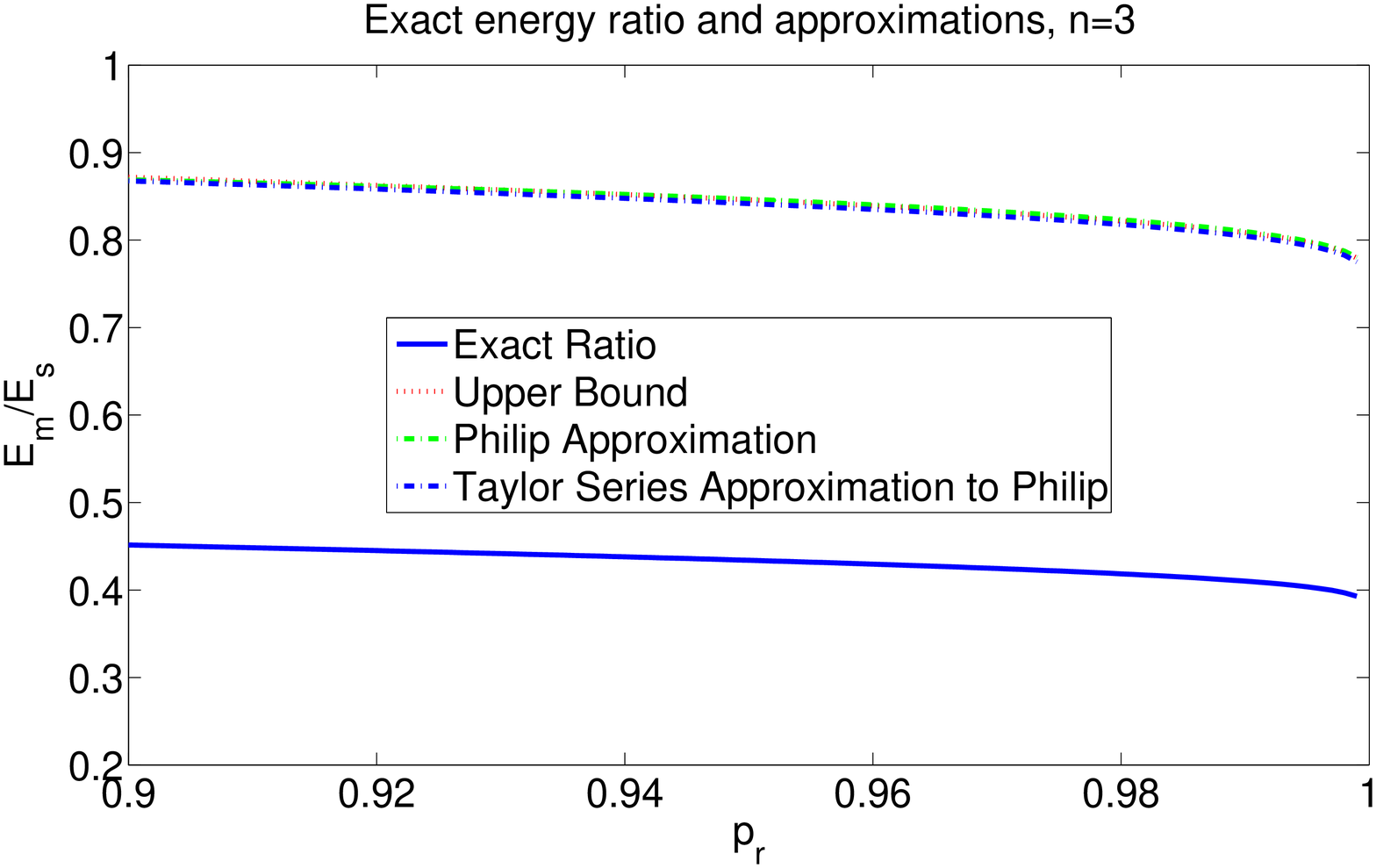}
\end{center}
\caption{Long-hop vs. short-hop energy for $n = 3$ hops.}
\label{sublinear-n3}
\end{figure}

Based on the tightness of the approximations that we have employed, as further evidenced by Fig. \ref{sublinear-n3}, we conclude that $f(p_r)$ is monotone decreasing for $p_r\in (0.9,1)$.  Note that $f(0.9) < \log(3)$, so we conclude that $k_m-k_s+1 < \log(3)$ for $p_r\in (0.9,1)$, establishing that short-hops are more energy-efficient than long-hops in this case for a deterministic MIMO line network.

\section{Computations for Example \ref{exp-2d-network}}\label{proof-exp-2}
Consider a random 2-D network with the parameters from Example \ref{exp-2d-network}.  Then, the ratio $E_A/E_B$ can be simplified as
\begin{equation}
\frac{E_A}{E_B} = \frac{1}{n}\cdot\frac{48n}{(48-\pi^2)n+\pi^2}\cdot\frac{2^{k_m}-1}{2^{k_s}-1}.
\end{equation}
Note that for $n = 3$, $(1/n)(48n/((48-\pi^2)n+\pi^2))\approx 0.386$.  Thus, if we can show that
\begin{equation}\label{short-better-2}
\begin{array}{lll}
\frac{2^{k_m}-1}{2^{k_s}-1} & < & \frac{1}{0.386} \\
& \approx & 2.59
\end{array}
\end{equation}
then short-hops will be more energy-efficient than long-hops in this case.

In particular, consider the following bounds
\begin{equation}\label{mystery-bound}
\begin{array}{lll}
\frac{2^{k_m}-1}{2^{k_s}-1} & \stackrel{(a)}{<} & \frac{2^{k_m}-1}{(3/2)\cdot 2^{k_s-1}} \\
& < & \frac{2}{3}\cdot 2^{k_m-k_s+1}.
\end{array}
\end{equation}
As shown in Section \ref{limit-line}, $2^{k_m-k_s+1} < n$ for $n = 3$.  Thus, if we can prove step (a) of \eqref{mystery-bound}, we will have established \eqref{short-better-2}.

From inspecting \eqref{mystery-bound}, we want to show that $2^{k_s}-1 > (3/2)\cdot 2^{k_s-1}$, which is equivalent to requiring that $(4/3) > 2^{k_s}/(2^{k_s}-1)$.  Note that the function $f(x) = 2^x/(2^x-1)$ is monotone decreasing.  In particular, as $p_r$ increases, $k_s$ increases based on the expressions in Section \ref{k-s-approx}.  Also, for $p_r = 0.9$ and $R = 4$, $2^{k_s}/(2^{k_s}-1) < 4/3$.

Thus, we have proved \eqref{mystery-bound} for $p_r\in (0.9,1)$ and established that short-hops are more energy-efficient than long-hops in this case for a random MIMO 2-D network.

\end{document}